\documentclass[conference]{IEEEtran}
\IEEEoverridecommandlockouts
\usepackage{cite}
\usepackage{amsmath,amssymb,amsfonts}
\usepackage{algorithmic}
\usepackage{graphicx}
\usepackage{textcomp}
\usepackage{xcolor}
\usepackage{float}
\usepackage[hidelinks]{hyperref}
\usepackage{url}
\usepackage{orcidlink}
\usepackage{eso-pic}
\def\BibTeX{{\rm B\kern-.05em{\sc i\kern-.025em b}\kern-.08em
    T\kern-.1667em\lower.7ex\hbox{E}\kern-.125emX}}

\usepackage{bbm} \usepackage{bm} 
\usepackage[all]{xy}

\newcommand{\IN}{\mathbb{N}}						\newcommand{\IR}{\mathbb{R}}						       		\newcommand{\ball}{\mathcal{B}}		              
\renewcommand{\epsilon}{\varepsilon}				\renewcommand{\theta}{\vartheta}                  	\renewcommand{\phi}{\varphi}						
\newcommand{\set}[2][]{\left\{ #2 
                                                \ifthenelse{\equal{#1}{}}{}{: #1}
                        \right\}}				        					\newcommand{\norm}[1]{\left\lVert #1 \right\rVert}		 					

\newcommand{\signatur}[3]{#1:#2\!\rightarrow#3}

\newcommand{\bxi}{ {\boldsymbol{x}_n} }

\newcommand{\bgam}{ {\boldsymbol\gamma} }

\newcommand{\numC}{N_c} \newcommand{\numP}{N_p} 
\usepackage[exponent-product = \cdot,separate-uncertainty=true]{siunitx}
\DeclareSIUnit{\mup}{\text{$\mu_0$}}
\DeclareSIUnit{\Mbps}{\text{Mbps}}

\usepackage{tikz}
\usepackage{tikz-3dplot}
\usepackage{pgfplots}
\usepackage{pgfplotstable}

\pgfplotsset{compat=1.8}

\usetikzlibrary{shapes,shapes.multipart,positioning,shapes.arrows, arrows.meta, decorations.pathmorphing, calc,fadings, plotmarks}
\usepgfplotslibrary{units,statistics,groupplots}

\pgfplotsset{
    unit markings=slash space,
    /pgfplots/xbar/.style={
    /pgf/bar shift={-0.5*(\numplotsofactualtype*\pgfplotbarwidth + (\numplotsofactualtype-1)*#1) + (.5+\plotnumofactualtype)*\pgfplotbarwidth + \plotnumofactualtype*#1},
    },
}
\pgfplotsset{select coords between index/.style 2 args={
    x filter/.code={
        \ifnum\coordindex<#1\fi
        \ifnum\coordindex>#2\fi
    }
}}

\definecolor{ibilight}{RGB}{193,216,237}
\definecolor{ibidark}{RGB}{0,73,146}	\definecolor{uke2}{RGB}{170,156,143} 	\definecolor{uke3}{RGB}{87,87,86}		\definecolor{ukesec1}{RGB}{255,223,0}	\definecolor{ukesec2}{RGB}{239,123,5}	\definecolor{ukesec3}{RGB}{104,195,205}	\definecolor{ukesec4}{RGB}{138,189,36}	\definecolor{ukesec5}{RGB}{178,34,41}	\definecolor{tuhh}{RGB}{45,198,214}     \definecolor{ibidarkBG}{RGB}{227,229,242}   \definecolor{uke2BG}{RGB}{233,228,225} 	    \definecolor{uke3BG}{RGB}{230,231,232}	    \definecolor{ukesec1BG}{RGB}{255,243,190}   \definecolor{ukesec2BG}{RGB}{254,232,212}   \definecolor{ukesec3BG}{RGB}{222,241,241}   \definecolor{ukesec4BG}{RGB}{233,243,222}   \definecolor{ukesec5BG}{RGB}{244,230,225}   

\usepackage[capitalise]{cleveref} 
\makeatletter
\newif\iftikz@shading@path

\tikzset{
                        shading xsep/.store in=\tikz@pathshadingxsep,
    shading ysep/.store in=\tikz@pathshadingysep,
    shading sep/.style={shading xsep=#1, shading ysep=#1},
    shading sep=0.0cm,
}

\def\tikz@shadepath#1{                \iftikz@shading@path    \else        \tikz@shading@pathtrue                \pgfgetpath\tikz@currentshadingpath                \begingroup            \pgfsys@beginscope            \tikzset{#1}            \xdef\tikz@tmp{\noexpand\def\noexpand\tikz@pathshadingxsep{\tikz@pathshadingxsep}                \noexpand\def\noexpand\tikz@pathshadingysep{\tikz@pathshadingysep}}            \pgfsys@endscope        \endgroup
        \tikz@tmp                \pgfextract@process\pgf@shadingpath@southwest{\pgfpointadd{\pgfqpoint{\pgf@pathminx}{\pgf@pathminy}}            {\pgfpoint{-\tikz@pathshadingxsep}{-\tikz@pathshadingysep}}}        \pgfextract@process\pgf@shadingpath@northeast{\pgfpointadd{\pgfqpoint{\pgf@pathmaxx}{\pgf@pathmaxy}}            {\pgfpoint{\tikz@pathshadingxsep}{\tikz@pathshadingysep}}}                \pgfsetpath\pgfutil@empty                \let\tikz@options@saved=\tikz@options        \let\tikz@mode@saved=\tikz@mode        \let\tikz@options=\pgfutil@empty        \let\tikz@mode=\pgfutil@empty                \tikz@addoption{            \pgfinterruptpath            \pgfinterruptpicture                \begin{tikzfadingfrompicture}[name=.]
                \pgfscope                    \tikzset{shade path/.style=}                    \path \pgfextra{                                                                        \pgfsetpath\tikz@currentshadingpath                                                \pgf@shadingpath@southwest
                        \expandafter\pgf@protocolsizes{\the\pgf@x}{\the\pgf@y}                        \pgf@shadingpath@northeast                        \expandafter\pgf@protocolsizes{\the\pgf@x}{\the\pgf@y}                                                \let\tikz@options=\tikz@options@saved                        \let\tikz@mode=\tikz@mode@saved                    };
                                        \xdef\pgf@shadingboundingbox@southwest{\noexpand\pgfqpoint{\the\pgf@picminx}{\the\pgf@picminy}}                    \xdef\pgf@shadingboundingbox@northeast{\noexpand\pgfqpoint{\the\pgf@picmaxx}{\the\pgf@picmaxy}}                    \endpgfscope
                \end{tikzfadingfrompicture}            \endpgfinterruptpicture            \endpgfinterruptpath                        \pgftransformreset            \pgfpathrectanglecorners{\pgf@shadingboundingbox@southwest}{\pgf@shadingboundingbox@northeast}                                    \let\tikz@path@picture=\pgfutil@empty            \tikz@mode@fillfalse            \tikz@mode@drawfalse                        \tikz@mode@doublefalse            \tikz@mode@clipfalse            \tikz@mode@boundaryfalse            \tikz@mode@fade@pathfalse            \tikz@mode@fade@scopefalse                        \tikzset{#1}            \tikz@mode                        \def\tikz@path@fading{.}            \tikz@mode@fade@pathtrue            \tikz@fade@adjustfalse                        \pgfpointscale{0.5}{\pgfpointadd{\pgf@shadingboundingbox@southwest}{\pgf@shadingboundingbox@northeast}}            \edef\tikz@fade@transform{shift={(\the\pgf@x,\the\pgf@y)}}            \pgfsetfading{\tikz@path@fading}{\tikz@do@fade@transform}            \tikz@mode@fade@pathfalse        }    \fi}
\tikzset{
    shade path/.code={        \tikz@addmode{\tikz@shadepath{#1}}    }
}
\makeatother 
\pgfplotsset{
colormap/inferno/.style={    /pgfplots/colormap={inferno}{      rgb=(0.001462, 0.000466, 0.013866)
      rgb=(0.037668, 0.025921, 0.132232)
      rgb=(0.116656, 0.047574, 0.272321)
      rgb=(0.217949, 0.036615, 0.383522)
      rgb=(0.316282, 0.053490, 0.425116)
      rgb=(0.410113, 0.087896, 0.433098)
      rgb=(0.503493, 0.121575, 0.423356)
      rgb=(0.596940, 0.154848, 0.398125)
      rgb=(0.688653, 0.192239, 0.357603)
      rgb=(0.775059, 0.239667, 0.303526)
      rgb=(0.851384, 0.302260, 0.239636)
      rgb=(0.912966, 0.381636, 0.169755)
      rgb=(0.956852, 0.475356, 0.094695)
      rgb=(0.981895, 0.579392, 0.026250)
      rgb=(0.987464, 0.690366, 0.079990)
      rgb=(0.973088, 0.805409, 0.216877)
      rgb=(0.947594, 0.917399, 0.410665)
      rgb=(0.988362, 0.998364, 0.644924)
    }
  },
    colormap/vik/.style={    /pgfplots/colormap={vik}{      rgb=(0.001328,0.069836,0.379529)
      rgb=(0.008947,0.163152,0.438691)
      rgb=(0.010254,0.253516,0.49702)
      rgb=(0.03198,0.349543,0.558906)
      rgb=(0.146607,0.458643,0.631289)
      rgb=(0.317174,0.578997,0.71295)
      rgb=(0.503691,0.69791,0.79398)
      rgb=(0.690457,0.811742,0.871059)
      rgb=(0.868874,0.900206,0.914812)
      rgb=(0.933752,0.858522,0.815421)
      rgb=(0.89269,0.743148,0.657086)
      rgb=(0.844258,0.629296,0.505146)
      rgb=(0.798475,0.52238,0.362835)
      rgb=(0.753316,0.419216,0.22819)
      rgb=(0.684653,0.299264,0.095633)
      rgb=(0.562268,0.168171,0.022523)
      rgb=(0.449521,0.078741,0.025203)
      rgb=(0.350423,6.1e-5,0.030499)
    }
  },
    colormap/lipari/.style={    /pgfplots/colormap={lipari}{      rgb=(0.0114, 0.0732, 0.1483)
      rgb=(0.0361, 0.161, 0.2734)
      rgb=(0.1054, 0.2519, 0.394)
      rgb=(0.2335, 0.3295, 0.4701)
      rgb=(0.3415, 0.3623, 0.478)
      rgb=(0.4193, 0.3706, 0.4615)
      rgb=(0.5002, 0.3756, 0.4404)
      rgb=(0.5923, 0.381, 0.4164)
      rgb=(0.6969, 0.3888, 0.3902)
      rgb=(0.811, 0.4119, 0.3704)
      rgb=(0.8985, 0.4821, 0.3842)
      rgb=(0.9158, 0.5797, 0.4364)
      rgb=(0.9014, 0.6653, 0.4992)
      rgb=(0.9034, 0.7523, 0.5845)
      rgb=(0.9384, 0.8531, 0.7098)
      rgb=(0.9923, 0.959, 0.8566)
    }
  },
    colormap/bluewhite/.style={    /pgfplots/colormap={bluewhite}{      color=(ibidark) 
      color=(white)
    }
  },
    colormap/whiteblue/.style={    /pgfplots/colormap={whiteblue}{      color=(white) 
      color=(ibidark)
    }
  },
  }

\begin{document}
\AddToShipoutPictureBG*{\AtPageUpperLeft{\raisebox{-1.2cm}{\makebox[\paperwidth]{%
  \parbox{0.85\paperwidth}{\footnotesize\centering © 2026 IEEE. Personal use of this material is permitted. Permission from IEEE must be obtained for all other uses, in any current or future media, including reprinting/republishing this material for advertising or promotional purposes, creating new collective works, for resale or redistribution to servers or lists, or reuse of any copyrighted component of this work in other works.}}}}}

\title{Maxwell-Consistent 3-D Magnetic Field Camera using Scalar-Potential-Based Reconstruction\\
{}
}

\author{\IEEEauthorblockN{Paul Jürß\IEEEauthorrefmark{1}\IEEEauthorrefmark{2}\orcidlink{0000-0002-3865-4603},
Marija Boberg\IEEEauthorrefmark{1}\IEEEauthorrefmark{2}\orcidlink{0000-0003-3419-7481},
Philip Suskin\IEEEauthorrefmark{1}\IEEEauthorrefmark{2}\orcidlink{0009-0009-0195-4491},
Tobias Knopp\IEEEauthorrefmark{1}\IEEEauthorrefmark{2}\IEEEauthorrefmark{3}\orcidlink{0000-0002-1589-8517}
Fynn Foerger\IEEEauthorrefmark{1}\IEEEauthorrefmark{2}\IEEEauthorrefmark{3}\orcidlink{0000-0002-3865-4603}, 
}
\IEEEauthorblockA{
Email: paul.juerss@tuhh.de\\
\IEEEauthorrefmark{1}Section for Biomedical Imaging, University Medical Center Hamburg-Eppendorf, Hamburg, Germany\\
\IEEEauthorrefmark{2}Institute for Biomedical Imaging, Hamburg University of Technology, Hamburg, Germany\\
\IEEEauthorrefmark{3}Fraunhofer Research Institution for Individualized Medical Technology and Engineering IMTE, Lübeck, Germany
}}

\maketitle
\begin{abstract}
Various applications require accurate 3-D characterization of magnetic fields with high spatial and temporal resolution. 
Building on our previously introduced spherical magnetic field camera employing 86 three-axis Hall sensors arranged on a spherical surface, we present a novel magnetic-field reconstruction approach based on the magnetic scalar potential and its spherical harmonic expansion. Unlike the previous reconstruction method that expands each vector component separately, our scalar-potential-based formulation is inherently consistent with Maxwell’s equations for source-free fields by design, yielding curl-free and divergence-free field models. Using a simulated magnetic gradient field generated by two permanent magnets, we demonstrate that the proposed method achieves higher reconstruction accuracy, improved robustness to sensor noise and failure, and substantially reduced non-physical divergence and curl compared to the component-wise baseline. Furthermore, the scalar-potential approach attains superior performance with fewer coefficients, indicating a more efficient use of measurement data allowing higher-degree polynomial representations. 
\end{abstract}


\section{Introduction}
Accurate knowledge of 3-D magnetic field distributions is essential in systems that rely on well-defined magnetic field configurations, including tomographic imaging~\cite{dietrich_field_2016}, magnetic tracking~\cite{son_simultaneous_2019}, and magnetic manipulation~\cite{yang_magnetic_2020}. In these applications, the spatial and temporal field distribution directly affects image quality, localization accuracy, or actuation performance.

Magnetic field cameras provide a practical way to access this information by acquiring magnetic field data at multiple spatial positions and converting these measurements into a field model in the region of interest. Such systems can reduce measurement time, enable repeated or dynamic field characterization, and provide feedback for calibration, monitoring, or control~\cite{Barmet2009FieldCamera,dietrich_field_2016,suksmono_magnetic_2021}.

Recently, we introduced a spherical 3-D magnetic field camera based on 86 three-axis Hall magnetometers placed according to a spherical $12$-design~\cite{FieldCam2026}. The system measures vector-valued magnetic field data at the boundary of a spherical volume with a \qty{90}{\milli\meter} diameter at an acquisition rate of \SI{10}{\Hz} to reconstruct the magnetic field inside the sphere.

While the original reconstruction framework expands the three magnetic field components separately in solid spherical harmonics, source-free magnetic fields also admit a scalar-potential formulation~\cite{Jackson2009Classical}. In this work, we introduce a scalar-potential reconstruction mode for the spherical magnetic field camera, in which a single harmonic potential is estimated from the surface measurements and the volumetric magnetic flux density is obtained by analytical differentiation. This representation enforces the curl-free and divergence-free structure of source-free magnetic fields by design and therefore provides a Maxwell-consistent field model. In addition, fields with more complex profiles can be represented, since this method can produce a higher-order underlying field model than the original component-wise framework, even though the same measurement data is used.

\section{Methods}
\begin{figure}
    \centering
    \includegraphics{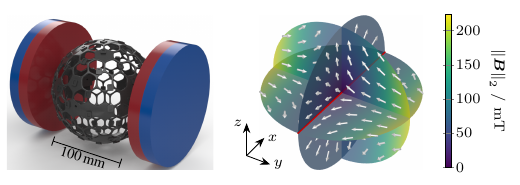}
    \caption{CAD rendering of the simulated assembly (left). Two permanent magnets with opposite polarities, spaced \qty{100}{\milli\meter} apart, generate a magnetic gradient field with a central field-free point. The 3D-printed mount for the Hall sensors of the field camera is shown in the center. On the right, the magnetic gradient field inside the field camera is shown. The red line indicates the axis along which the results in \autoref{fig:vplot} are shown.}
    \label{fig:sph_render}
\end{figure}

\subsection{Magnetic Scalar Potential}
Considering a magnetic field $\signatur{\bm B}{\IR^3}{\IR^3}$ inside a simply connected region without currents, it can be expressed in terms of the magnetic scalar potential $\signatur{\psi}{\IR^3}{\IR}$ as $\bm B\!=\!-\nabla\psi.$
With Gauss' law for magnetism, it follows that $\Delta\psi=0$. Thus, the magnetic scalar potential can be written as a spherical harmonic expansion
\begin{equation}\label{eq:sp_expansion}
    \psi(\bm x) = \sum_{\ell=0}^\infty\sum_{m=-\ell}^\ell \gamma_{\ell, m}Z_\ell^m(\bm x),
\end{equation}
where $Z_\ell^m:\IR^3\to\IR$ denote the normalized real solid spherical harmonics~\cite{Boberg2025Unique} and $\gamma_{\ell,m}\in\IR$ denote the corresponding spherical harmonic coefficients~\cite{Jackson2009Classical}. In practice, this expansion is truncated at a maximum polynomial degree $L+1\in\IN$. Consequently, the magnetic field itself can be represented in a similar way
\begin{equation}\label{eq:SHE_B}
    \bm B(\bm x) \approx -\sum_{\ell=1}^{L+1}\sum_{m=-\ell}^\ell \gamma_{{\ell}, m}\nabla Z_{\ell}^m(\bm x)
\end{equation} 
with $\numC = (L+2)^2 - 1$ coefficients.  For a specific position $\bxi\in\IR^3$, the expansion can be compactly written as a system of equations $\mathbf B_{\bxi} = \mathbf W_{\bxi} \mathbf c$ with

\begin{align*}
    \mathbf B_{\bxi} &= \bm B(\bxi) \in \IR^{3},\quad \mathbf c = \left(\gamma_{\ell,m}\right)_{\substack{\ell = 1,\dots,L+1 \\ m = -\ell,\dots,\ell}} \in \IR^{\numC},\\
    \mathbf W_{\bxi} &= -\left( \nabla Z_\ell^m(\bxi) \right)_{\substack{\ell = 1,\dots,L+1 \\ m = -\ell,\dots,\ell}} \in \IR^{3\times \numC}.
\end{align*}

In general, the coefficients $\gamma_{\ell,m}$ can be uniquely determined by imposing Dirichlet boundary conditions on the surface of the ball $\partial\ball_R = \set{\bm x \in \IR^3 : \norm{\bm x}_2 = R}$. However, the magnetic scalar potential cannot be measured directly, only its gradient, i.e., the magnetic field.
Following the design of our field camera~\cite{FieldCam2026}, the coefficients are reconstructed from the measured vector-valued magnetic field on $\partial\ball_R$.
Assuming field values at $\numP\in\IN$ discrete positions $\bm x_n \in \partial\ball_R$, $n\in\set{1,\dots,\numP}$, on the boundary are provided, the coefficients can be calculated using the least squares approach 
\begin{align}\label{eq:leastSquares}
    \bgam = \underset{\mathbf c \in \IR^{\numC}}{\text{argmin}}
    \norm{
    \mathbf W \mathbf c - \mathbf B
    }_2^2.
\end{align}
with the stacked matrices
\begin{align*}
        \mathbf B &= \left(\mathbf B^\top_{\bm x_1},\dots, \mathbf B^\top_{\bm x_N}\right)^\top \in \IR^{3\numP}~\text{and}\\
    \mathbf W &= \left( \mathbf W^\top_{\bm x_1},\dots, \mathbf W^\top_{\bm x_N} \right)^\top \in \IR^{3\numP\times \numC}.
\end{align*}

\subsection{Experiments}
The proposed method was evaluated using a simulated gradient field, since systematic sensor imperfections can introduce non-physical divergence and curl, preventing experimental measurements from serving as an exact physical ground truth. The simulation therefore enabled reconstruction accuracy and consistency with Maxwell's equations to be assessed under controlled conditions and allowed a direct comparison between the baseline component-wise spherical harmonics representation and the proposed scalar-potential-based formulation.

For the simulated field, finite-element analysis was performed using the AC/DC Module in \textsc{COMSOL Multiphysics} \cite{COMSOL63}. As shown in \autoref{fig:sph_render}, the model consisted of two cylindrical, axially magnetized N52-grade NdFeB permanent magnets with a radius of \SI{50}{\mm} and a thickness of \SI{20}{\mm}. The magnets were separated by \SI{100}{\mm} and magnetized in opposite directions along their common symmetry axis.

The simulated field was evaluated at the \num{86} positions of a spherical $12$-design~\cite{Beentjes2015quadrature} with radius \qty{45}{\milli\meter}, corresponding to the sensor positions of the field camera~\cite{FieldCam2026}, and on a $31\times31\times31$ Cartesian grid with a side length of \qty{90}{\milli\meter} both centered at the origin. According to the specifications of the Hall sensor employed in the field camera (TMAG5273, Texas Instruments), white Gaussian noise with a standard deviation (SD) of \qty{145}{\micro\tesla} was added to the field values.

The baseline method featuring a component-wise expansion~\cite{FieldCam2026} and the proposed method were compared for different polynomial degrees $L$ of the field expansion in terms of the mean relative absolute error (MRAE) of the reconstructed magnetic field relative to the ground truth grid values within the measurement volume, excluding field values below the noise threshold of \qty{145}{\micro\tesla}. In addition, the robustness to single sensor failures was investigated by retroactively removing each sensor individually before reconstruction. The robustness to noise was investigated for \num{100} different noise levels with a SD logarithmically spaced between \qty{e-6}{\tesla} and \qty{e-3}{\micro\tesla}. To assess the physical consistency of the reconstructed fields, non-physical divergence and curl were quantified for both methods. For the scalar-potential-based formulation, residual values may still arise from numerical integration and discretization.

\subsection{Evaluation}
Non-physical divergence was quantified by comparing the magnetic flux entering and leaving the spherical measurement volume. The inward and outward fluxes were computed as

\begin{equation*}
\Phi_\mathrm{in}\!=\!\int_{\partial\ball_R}\!\!(-\bm B\!\cdot\!\bm n)_+\,\mathrm dA,\quad
\Phi_\mathrm{out}\!=\!\int_{\partial\ball_R}\!\!(\bm B\!\cdot\!\bm n)_+\,\mathrm dA,
\end{equation*}
where $\bm{n}$ denotes the surface normal and $x_+\!=\!\max\{x,0\}$. The relative flux imbalance can then be defined as
\begin{equation}
    \varepsilon_\mathrm{div}
    =
    \frac{\Phi_\mathrm{out}-\Phi_\mathrm{in}}{\Phi_\mathrm{in}} .
\end{equation}

Curl-related inconsistencies were expressed as apparent currents through the three central coordinate planes that would be required to generate the rotational components observed in the reconstructed field. For a cross-sectional area $A_i$ of the spherical measurement volume with normal direction $\bm{e}_i$, the mean apparent current was computed as

\begin{equation}
    \bar I_{\mathrm{app}}=\frac{1}{3\mu_0}\sum_{i \in \{x,y,z\}}\int_{A_i}\left(\nabla\times\bm{B}\right)\cdot\bm{e}_i\,\mathrm{d}A.
\end{equation}

\section{Results}
\begin{figure}
    \centering
    \includegraphics{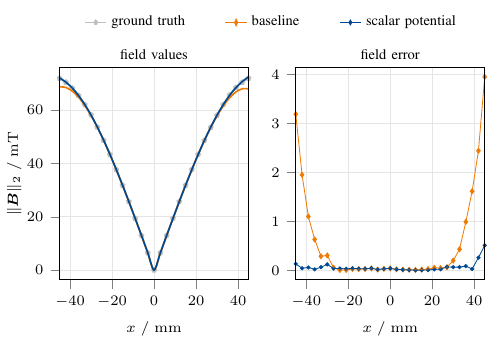}
    \caption{Comparison of field values (left) and absolute difference (right) to the ground truth along the $x$-axis of both scalar potential with $L=10$ (blue) and baseline $L=6$ (orange) with respect to the ground truth values (gray).}
    \label{fig:vplot}
\end{figure}

The qualitative results of both methods along the $x$-axis (see \autoref{fig:sph_render}) with a comparable number of coefficients are shown in \autoref{fig:vplot}. The spherical $12$-design allows exact determination of spherical harmonic coefficients up to degree $6$, which corresponds to \num{147} coefficients. In contrast, with a similar number of coefficients, the more efficient use of them in the scalar-potential-based method enables an expansion up to $L=10$ (\num{143} coefficients). Across the region of interest, the baseline achieved an MRAE of \qty{0.7}{\percent} and the proposed method a value of \qty{0.1}{\percent}.

\begin{figure}
    \centering
    \includegraphics{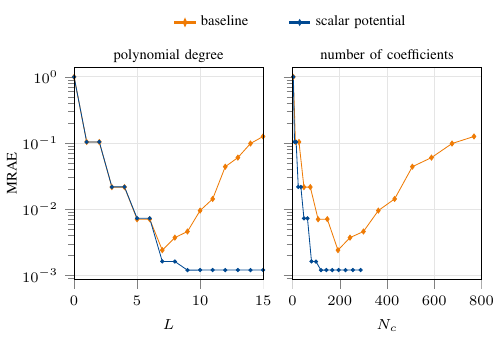}
    \caption{Comparison of proposed method (blue) with the baseline (orange) in terms of MRAE on the ground truth for different polynomial order $L$ (left) and number of coefficients $\numC$ (right).}
    \label{fig:convergence}
\end{figure}

The MRAE of both methods for different levels of truncation is shown in \autoref{fig:convergence}. The baseline achieved a minimal MRAE of \qty{0.2}{\percent} with $L=7$. Beyond $L=7$, the baseline error increases. For $L\leq6$ both baseline and proposed method show comparable performance. However, for $L>6$ the proposed method shows superior performance converging at an MRAE of approximately \qty{0.1}{\percent}, which is first achieved for $L=9$. The difference in performance becomes even more apparent when comparing both methods in terms of the number of coefficients as the baseline requires a set of spherical harmonic coefficients for each component of the magnetic field ($\numC\!=\!3(L\!+\!1)^2$) whereas the proposed method requires only a single set ($\numC\!=\!(L\!+\!2)^2\!-\!1$).
The same trend can be observed for the absolute difference. In the following, results will only be reported with $L\!=\!6$ for the baseline and $L\!=\!10$ for the proposed method.

Consistent with the results presented above, the proposed method demonstrates superior robustness to the failure of a single sensor. Across all \num{86} sensors, the MRAE of the scalar potential estimation was \qty{0.116+-0.005}{\percent}. By comparison, the baseline approach exhibited a higher MRAE of \qty{0.74+-0.03}{\percent}. Similarly, the scalar-potential-based approach showed superior performance across a range of different noise levels with a mean difference between the MRAE values of both methods of \qty{0.58+-0.02}{\percent}.

Regarding physical consistency, the proposed method outperforms the baseline by several orders of magnitude. The relative flux imbalance $\epsilon_\mathrm{div}$ caused by divergence was approximately \num{-4e-4} for the baseline, whereas the proposed method achieved a near-zero value of \num{-6e-16}. Similarly, the mean current $\bar I_\mathrm{app}$ associated with non-zero curl was \qty{3e-4}{\ampere} for the baseline, but only \qty{8e-20}{\ampere} for the proposed method.

\section{Discussion and Conclusion}
The proposed method outperformed the baseline in reconstruction accuracy and robustness, especially when comparing them at similar coefficient counts. Although both approaches possess comparable representation capabilities at equivalent degrees $L$, the proposed approach utilizes data points more efficiently due to its intrinsically lower coefficient count per degree resulting in a higher degree of overdetermination. Consequently, it maintains a lower MRAE under noise and sensor dropout, highlighting its increased robustness compared to the baseline. The flux imbalance and apparent currents demonstrate that the baseline component-wise reconstruction can produce Maxwell-inconsistent fields, since the three field components are treated independently. This issue may be amplified in practical measurements by sensor imperfections such as cross-axis coupling, whereas the scalar-potential-based formulation avoids such inconsistencies by design.

Although the baseline guarantees exact field recovery up to degree $6$ in noise-free conditions, this exactness relies on the strict assumption that the measured field belongs to the corresponding finite-dimensional function space. Because field structures are rarely known a priori in practice, this theoretical exactness offers limited practical benefit. Consequently, the proposed field reconstruction is treated as an approximation problem from discrete, noisy sensor data. Although the baseline approach can also be extended using a least squares formulation, empirical results demonstrate that it is less accurate than the proposed method. This method simplifies accounting for different sensor position distributions or shifts in a sensor's active regions for the three field components, respectively. However, the spherical $t$-design remains a well-suited sensor arrangement, as its nearly uniform distribution avoids directional bias.

One potential advantage of using the scalar potential is that it can be determined solely from the normal component of the magnetic field~\cite{Blakely2009Potential}, enabling the use of single-axis sensors. Consequently, this broadens the range of suitable sensor technologies and may thereby enable application-specific improvements such as higher dynamic range, faster measurement speed, or reduced noise.

In conclusion, we investigated the feasibility of using the scalar potential in a previously proposed field camera to represent the magnetic field. This new reconstruction procedure has shown superior performance and robustness compared to the previous component-wise spherical harmonic-based approach, while having an improved efficiency with respect to the number of coefficients. Future work will have to validate the presented findings experimentally with a more diverse set of field  and investigate the potential of regularization in \eqref{eq:leastSquares}.

\bibliographystyle{IEEEtran}

\bibliography{ref}

\end{document}